\documentclass[twocolumn,twoside]{IEEEtran}
\usepackage{amsmath,amssymb,amsthm,wasysym,epsfig,color,subfigure,graphicx}
\usepackage{enumerate,url,algpseudocode,algorithm,balance,url,bm,nicefrac,caption,subfloat}
\usepackage[table]{xcolor}
\usepackage{stfloats}
\usepackage{graphicx}
\usepackage{subfigure}

\newtheorem{proposition}{Proposition}

\theoremstyle{remark}\newtheorem{remark}{Remark}
\allowdisplaybreaks

\allowdisplaybreaks

\begin{document}
	%\bstctlcite{IEEEexample:BSTcontrol}
%	
%	\title{Physics-driven Graph neural  Networks for Robust
%       Power System State Estimation}	

%\title{Go Graph for Power System State Estimation}	
%\title{Graph CNNs for Robust PSSE}
%\title{Robust PSSE With Graph Neural Network Priors}
\title{Gauss-Newton Unrolled Neural Networks and Data-\\driven 
Priors for Regularized PSSE with Robustness}
%\title{Robust PSSE Using Graph Neural Networks 
%	for Data-driven and Topology-aware Priors }
%\title{Data-driven Priors for Robust Power\\ System State Estimation}

	\author{
		Qiuling Yang,~Alireza Sadeghi,~Gang Wang, 
Georgios B. Giannakis,~\IEEEmembership{Fellow,~IEEE}, and Jian Sun%,  \IEEEmembership{Member,~IEEE}
		
		\thanks{The work of Q. Yang and J. Sun was supported in part by NSFC  
			Grants 61522303, 61720106011, 61621063, and U1613225.
			Q. Yang was also supported by the China Scholarship Council. 
			The work of A. Sadeghi, G. Wang, and G. B. Giannakis was supported by NSF	grants 1711471 and 1901134.			
			Q. Yang and J. Sun are with the State Key Lab of Intelligent Control and Decision of Complex Systems, School of Automation, 
			Beijing Institute of Technology, Beijing 100081, China (e-mail: yang6726@umn.edu, sunjian@bit.edu.cn). 	
			 A. Sadeghi, G. Wang, and G. B. Giannakis are with the Department of Electrical and Computer Engineering, University of Minnesota, Minneapolis, MN 55455, USA
			(e-mail:  sadeghi@umn.edu, gangwang@umn.edu,  georgios@umn.edu).
		}
	}
	
%	\markboth{IEEE TRANSACTIONS ON SMART GRID (Re-Revised, \today)}{}
	\maketitle
	
\begin{abstract}
Renewable energy sources, elastic loads, and purposeful manipulation of meter readings challenge the monitoring and control of today's power systems (PS). In this context, fast and robust state estimation (SE) is timely and of major importance to maintain a comprehensive view of the system in real time. Conventional PSSE solvers typically entail minimizing a nonlinear and nonconvex least-squares cost using e.g., the Gauss-Newton method. Those iterative solvers however, are sensitive to initialization, and may converge to local minima. To overcome these hurdles, the present paper adapts and leverages recent advances on image denoising to introduce a PSSE approach with a regularizer capturing a deep neural network (DNN) prior. For the resultant regularized PSSE objective, a ``Gauss-Newton-type" alternating minimization solver is developed first. To accommodate real-time monitoring, a novel end-to-end DNN is constructed subsequently by unrolling the proposed alternating minimization solver. The deep PSSE architecture can further account for the power network topology through a graph neural network (GNN) based prior. To further endow the physics-based DNN with robustness against bad data, an adversarial DNN training method is put forth. Numerical tests using real load data on the IEEE $118$-bus benchmark system showcase the improved estimation and robustness performance of the proposed scheme compared with several state-of-the-art alternatives. 
\end{abstract}
	

\section{Introduction} \label{Sec:Intro}
In today's smart grid, reliability and accuracy of state estimation 
are central for several system control and optimization tasks, including optimal power flow, unit commitment, economic dispatch, and contingency analysis \cite{pssebook2004}.
However, frequent and sizable state variable fluctuations caused by fast variations of renewable generation, increasing deployment of electric vehicles, and human-in-the-loop demand response incentives, are challenging these functions. 

As state variables are difficult to measure directly, the supervisory control and data acquisition (SCADA) system  offers abundant measurements, including voltage magnitudes, power flows, and power injections. Given SCADA measurements, the goal of PSSE is to retrieve the state variables, namely complex voltages at all buses \cite{pssebook2004}. PSSE is typically formulated as a (weighted) least-absolute-value (WLAV) or a  least-squares (WLS) problem, both of which can be underdetermined, and nonconvex in general \cite{wang2019overview}. 

To address these challenges, several efforts have been devoted. WLAV-based estimation for instance can be converted into a constrained optimization, for which a sequential linear programming solver was devised in \cite{jabr2003iteratively}, and improved (stochastic) proximal-linear solvers were developed in \cite{gang2019tsg}. On the other hand, focusing on the WLS criterion, the Gauss-Newton solver is widely employed in practice \cite{pssebook2004}. Unfortunately, due to the nonconvexity and quartic loss function, there are two challenges facing the Gauss-Newton solver: i) sensitivity to initialization; and ii) convergence is generally not guaranteed~\cite{zhu2014power}. Semidefinite programming approaches can mitigate these issues to some extent, at the price of rather heavy computational burden \cite{zhu2014power}. In a nutshell, the grand challenge of these methods, remains to develop fast and robust PSSE solvers attaining or approximating the global optimum.    
 
To bypass the nonconvex optimization hurdle in power system monitoring and control, recent works have focused on developing data- (and model-) driven  neural network (NN) solutions  \cite{barbeiro2014state,tps2012nnpsse,zhang2018real,liang2019tsgpsse,zamzam2019gcn, chen2020input, yang2019twotimerltsg,  hu2020physics, zamzam2019learning, tps20langtong,ostrometzky2019physics,tps2020dobbe}.
%A neural network (NN) was trained using historical data to obtain a `smart' initialization for Gauss-Newton iterations in \cite{zamzam2019datadriventps}.
Such NN-based PSSE solvers approximate the mapping from measurements to state variables based on a training set of measurement-state pairs generated using simulators or available from historical data \cite{liang2019tsgpsse}. However, existing NN architectures do not directly account for the power network topology. On the other hand, a common approach to tackling challenging ill-posed problems in image processing has been to regularize the loss function with suitable priors \cite{rudin1992nonlinear}. Popular priors include sparsity, total variation, and low rank \cite{bookregularization}. Recent efforts have also focused on data-driven priors that can be learned from exemplary data \cite{learningprior2013mri,schlemper2017deep, modl2018tmi}.  

Permeating the benefits of \cite{learningprior2013mri,schlemper2017deep} and \cite{modl2018tmi} to power systems, this paper advocates a deep (D) NN-based trainable prior for standard ill-posed PSSE, to promote physically meaningful PSSE solutions. To tackle the resulting regularized PSSE problem, an alternating minimization-based solver is first developed, having Gauss-Newton iterations as a critical algorithmic component. As with Gauss-Newton iterations, our solver requires inverting a matrix per iteration, thus incurring a heavy computational load that may discourage its use for real-time monitoring of large networks. To accommodate real-time operations and building on our previous works \cite{zhang2018real,liang2019tsgpsse}, we unroll this alternating minimization solver to construct a new DNN architecture, that we term Gauss-Newton unrolled neural networks  (GNU-NN) with deep priors. As the name suggests, our DNN model consists of a Gauss-Newton iteration as a basic building block, followed by a proximal step to account for the regularization term.  Upon incorporating a graph (G) NN-based prior, our model exploits the structure of the underlying power network. Different from \cite{liang2019tsgpsse}, our GNU-NN  method offers a systematic and flexible framework to incorporate prior information into standard PSSE tasks. 

In practice, measurements collected by the SCADA system may be severely corrupted due to e.g., parameter uncertainty, instrument mis-calibration, and unmonitored topology changes \cite{bad1971tpas, gang2019tsg}. As cyber-physical systems, power networks are also vulnerable to adversarial attacks 
%due to that they are designed without paying enough attention to security 
\cite{fairley2016spectrum,wu2019optimal,tac2020wu}, as asserted by the first hacker-caused Ukraine power blackout in $2015$ \cite{ukraincase2016analysis}. Furthermore,  it has recently been demonstrated that adversarial attacks can markedly deteriorate NNs' performance \cite{adv2016kurakin,miller2019adversarial}. Prompted by this, to endow our GNU-NN approach with \textit{robustness} against bad (even adversarial) data, we pursue a principled GNU-NN training method that relies on a distributionally robust optimization formulation. Numerical tests using the IEEE $118$-bus benchmark system corroborate the performance and robustness of the proposed scheme.

\emph{Paper outline.} 
Regarding the remainder of the paper, Section~\ref{sec:prob} introduces the power system model, and formally states the PSSE problem. Section \ref{sec:unro} presents a general framework for incorporating data-driven and topology-aware priors into PSSE, along with an alternating minimization solver for the resultant regularized PSSE. Section \ref{sec:adve} develops an adversarial training method to robustify GNU-GNN against bad data. Numerical tests using the IEEE $118$-bus test feeder are provided in Section \ref{sec:test}, with concluding remarks drawn in Section \ref{sec:conc}. 

\emph{Notation.} Lower- (upper-) case boldface letters denote column vectors (matrices), with the exception of vectors $\bm{V}$, $\bm{P}$ and $\bm{Q}$, and normal letters represent scalars. The $(i,j)$th entry, $i$-th row, and $j$-th column of matrix $\bm X$ are $[\bm X]_{i,j}$, $[\bm X]_{i:}$, and $[\bm X]_{:j}$, respectively. Calligraphic letters are reserved for sets except operators $\mathcal{I}$ and $\mathcal{P}$. Symbol $^{\top}$ stands for transposition; 
$\bf 0$ denotes all-zero vectors of suitable dimensions; and $\| \pmb x \|$ is the $l_2$-norm of vector $\pmb x$.
% Symbol 
%%$\bf 1$ denotes all-one vectors; 
%$\bf 0$ denotes all-zero vectors; $^{\top}$ stands for transposition; and $\| \pmb x \|$ is the $l_2$-norm of~$\pmb x$.

	%%%%%%%%%%%%%%%%%%%%%%%%%%%%%%%%%%%%%%%%%%%%%%%%%%%%%%%%%%%%%%%%%%%%%%%
	\section{Background and Problem Formulation}\label{sec:prob}

%\begin{figure}
%	\centering
%	\includegraphics[width =0.26 \textwidth]{powerflow.pdf}
%	\caption{Bus $n$ and $n'$ are connected with line impedance $y_{nn'}$.}
%	\label{fig:lineardiagram}
%\end{figure}	
	
Consider an electric grid comprising $N$ buses (nodes) with $E$ lines (edges) that can be modeled as a graph $\mathcal{G}:=(\mathcal{N},\mathcal{E}, \bm W)$, where the set $\mathcal{N}:=\{1,\ldots,N\}$ collects all buses, $\mathcal{E}:=\{(n, n')\} \subseteq \mathcal N \times \mathcal N$  all lines, and $\bm W \in \mathbb R^{N \times N}$ is a weight matrix with its $(n, n')$-th entry $[\bm W]_{nn'} = w_{nn'}$  modeling the impedance between buses $n$ and $n'$.
	 %impedance between buses $n$ and $n'$. 
%	 Among possible choices, we use a Gaussian kernel, that is $w_{nn'}=\exp(-k|y_{nn'}|^2)$, where $k$ is a scaling factor, and $y_{nn'}$ is the impedance between buses $n$ and $n'$. 
In particular, if $(n,n') \in \mathcal E$, then $[\bm W]_{nn'} = w_{nn'}$; and $[\bm W]_{nn'} = 0$ otherwise. 
%	whose buses are collected into $\mathcal{N}:=\{1,\ldots,N\}$, and lines into $\mathcal{E}:=\{(n, n')\} \subseteq \mathcal N \times \mathcal N$. 
	For each bus $n \in \mathcal N$, let $V_n := v^r_n + jv^i_n$ be its complex voltage with magnitude denoted by $|V_n|$, and $P_n + j Q_n$ its complex power injection. For reference, collect the voltage magnitudes, active and reactive power injections across all buses into the $N$-dimensional column vectors $|\bm  V|$, $\bm P$, and $\bm Q$, respectively.  
	 %For each line $(n, n') \in \mathcal L$, let $P^f_{nn'}$ denote the active power flow seen at the `from' end, and $P^t_{nn'}$ the active power flow seen at the `terminal' end, likewise for reactive power flow $Q^f_{nn'}$ (from) and $Q^t_{nn'}$ (terminal); 

\begin{figure*}[t!]
	\centering
	\includegraphics[width =0.9 \textwidth]{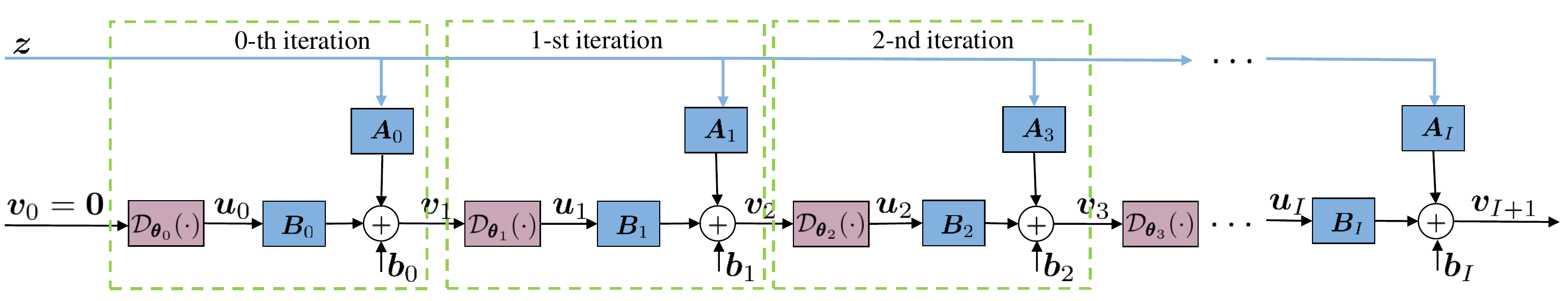}
	\caption{The structure of the proposed GNU-NN.}
	\label{fig:unrollGNN}
\end{figure*}

System state variables $\bm{v}:=[v^r_{1}~ v^i_{1} ~\dots  v^r_{N} ~ v^i_{N}]^\top \in \mathbb{R}^{2 N}$ can be represented by SCADA measurements, including voltage magnitudes, active and reactive power injections, as well as active and reactive power flows. Let $\mathcal S_V$, $ \mathcal S_P$, $\mathcal S_Q$, $\mathcal{E}_P$, and $\mathcal{E}_Q$ denote the sets of buses or lines where meters of corresponding type are installed. 
%  us signify the smart meter locations of corresponding types by $\mathcal S_V$, $ \mathcal S_P$ and $\mathcal S_Q$, where %at time slot $t$,  
%$M$ measurements are observed. 
% let $\mathcal C_V$, $ \mathcal C_P$ and $\mathcal C_Q$ $\mathcal C^f_P$, $\mathcal C^t_Q$, $\mathcal C^f_P$ and $\mathcal C^t_Q$ denote the smart meter locations of the corresponding type. 
For a compact representation, 
let us collect the measurements from all meters into 
$\bm{z}:=[\{|V_{n}|^2\}_{n \in \mathcal S_V}, \{P_{n}\}_{n \in \mathcal S_P}, \{Q_{n}\}_{n \in \mathcal S_Q}, \{P_{nn'}\}_{(n,n') \in \mathcal E_P},$ $ \{Q_{nn'}\}_{(n,n') \in \mathcal E_Q},]^\top\in \mathbb R^M$. 
%For brevity, the time index $t$ is dropped.
Moreover, the $m$-th entry of $\bm{z}:=\{z_{m}\}^M_{m=1}$, can be described by the following model 	
\begin{equation}\label{eq:measurement}
	z_{m} =  h_{m} (\bm v) + \epsilon_{m}, ~~ \forall m = 1, \ldots, M	 
	\end{equation}
where  $ h_m(\bm v)=\bm v ^\top \bm H_{m} \bm v $ for some symmetric measurement matrix $\bm H_m \in \mathbb R^{2N \times 2N}$, and $\epsilon_{m}$ captures the modeling error as well as the measurement noise. 

	The goal of PSSE is to recover the state vector $\bm v$ from measurements $\bm z$. Specifically, 
%	for independent and identically distributed noise $\{\epsilon_m\}_{m=1}^M$,
	 adopting the least-squares criterion and vectorizing the terms in \eqref{eq:measurement}, PSSE can be formulated as the following nonlinear least-squares (NLS) 
%	As the  $\ell_{1}$-based losses yield median-based estimators \cite{huber2011robust}, they handle gross errors in the measurements $\bm{z}$ in a relatively benign way. Motivated by this, we consider here  minimizing the $\ell_{1}$ loss of the residuals, which gives the so-called least-absolute-value estimate \cite{lav1982tpas}
%	\begin{equation}
%	\min _{\bm v \in \mathbb R^{2N}} \mathbb{E}_{\bm {z} \sim P_{0}}|\bm z - \bm v^{\top} \bm H \bm v|
%	\end{equation}	
%	The nominal distribution $P_{0}$ is typically unknown, but instead samples $\left\{z_{n}\right\}_{n=1}^{N} \sim P_{0}$ are given. Therefore, empirical loss minimization is used in practice, where $P_0$ is replaced with empirical distribution $\widehat{P}_{N}$, resulting in following reformulation
%	
%	\begin{equation}
%	\min _{\boldsymbol{\theta} \in \Theta} \mathbb{E}_{\boldsymbol{z} \sim \widehat{P}_{N}}[\ell(\boldsymbol{\theta} ; \boldsymbol{z})]+r(\boldsymbol{\theta})
%	\end{equation}
%	
%	where $\mathbb{E}_{\boldsymbol{z} \sim \widehat{P}_{N}}[\ell(\boldsymbol{\theta} ; \boldsymbol{z})]=N^{-1} \sum_{n=1}^{N} \ell\left(\boldsymbol{\theta} ; \boldsymbol{z}_{n}\right)$
	\begin{equation}\label{eq:exploss}
	 {\bm v}^\ast : = \arg \underset{\bm v \in \mathbb R^{2N}}{\min} ~ \|\bm{z} -  \bm h(\bm v) \|^2.
	\end{equation}
A number of algorithms have been developed for solving \eqref{eq:exploss}, including e.g., Gauss-Newton iterations \cite{pssebook2004}, and semidefinite programming-based solvers  \cite{gangzamzam2018cramer, zhu2014power, jin2018power,lan2019fast}. Starting from an initial $\bm v_{0}$, most of these schemes (the former two) iteratively implement a mapping from $\bm v_i$ to $\bm v_{i+1}$, in order to generate a sequence of iterates that hopefully converges to $\bm v^{\ast}$ or some point nearby. In the ensuing subsection, we will focus on the `workhorse' Gauss-Newton PSSE solver. 
%$\Gamma\left(x^{k}, A, y, \theta\right) \rightarrow x^{k+1}$ is 

\subsection{Gauss-Newton Iterations}
The Gauss-Newton method is the most commonly used one for minimizing NLS \cite[Sec. 1.5.1]{nolinear1999book}. It relies on Taylor's expansion to linearize the function $\bm h(\bm{v})$. Specifically, at a given point $\bm{v}_{i}$, it linearly approximates  
\begin{equation}
\label{eq:lin}
	\tilde{\bm{h}}(\bm{v}, \bm{v}_{i})\approx \bm{h}(\bm{v}_{i})+\bm{J}_{i}(\bm{v}-\bm{v}_{i})
\end{equation}
where $\bm{J}_i := \nabla \bm{h}\left(\bm{v}_i\right)$ is the $M \times 2N$ 
Jacobian of $\bm h$ evaluated at $\bm v_i$, with $[\bm{J}_i]_{m,n} :=\partial \bm{h}_{m} / \partial \bm v_{n}$.
%; see Wirtinger derivative \cite{kreutz2009complex} for details. %\cite{T. Lipp and S. Boyd, \Variations and extension of the convex{concaveprocedure," Optim. Eng., vol. 17, no. 2, pp. 263{287, Jun. 2016.}
Subsequently, the Gauss-Newton method approximates the nonlinear term $\bm{h}(\bm{v})$ in \eqref{eq:exploss} via \eqref{eq:lin}, and finds the next iterate as its minimizer; that is,
\begin{equation}\label{eq:gaussnewton}
\bm v_{i+1}= \arg \min_{\bm v}~ \left\| \bm{z} - \bm h(\bm v_i)- \bm J_i(\bm v - \bm v_i)\right\|^2.
\end{equation}
Clearly, the per-iteration subproblem \eqref{eq:gaussnewton} is convex quadratic. 
%The optimal $\bm v^{\ast}$ can be found by means of standard convex programming methods. 
If matrix $\bm{J}_i^{\top} \bm{J}_i$ is invertible, the iterate $\bm v_i$ can be updated in closed-form as
\begin{equation}\label{eq:closeform}
	\bm{v}_{i+1}=\bm{v}_{i}+\big(\bm{J}_{i}^{\top} \bm{J}_{i}\big)^{-1}\bm{J}_{i}^{\top}(\bm{z}-\bm{h}(\bm{v}_{i}))
\end{equation}
until some stopping criterion is satisfied. In practice however, due to the matrix inversion, the Gauss-Newton method becomes computationally expensive; it is also sensitive to initialization, and in certain cases it can even diverge. These limitations discourage its use for real-time monitoring of large-scale networks. To address these limitations, instead of solving every PSSE instance (corresponding to having a new set of measurements in $\bm{z}$) with repeated iterations, an end-to-end approach based on DNNs is pursued next.

%%%%%%%%%%%%%%%%%%%%%%%%%%%%%%%%%%%%
\section{Unrolled Gauss-Newton with Deep Priors}\label{sec:unro}
As mentioned earlier, PSSE can be underdetermined and thus ill posed due to e.g., lack of observability. To cope with such a challenge, this section puts forth a flexible topology-aware prior that can be incorporated as a regularizer of the PSSE cost function in~\eqref{eq:exploss}. To solve the resultant regularized PSSE, an alternating minimization-based solver is developed. Subsequently, an end-to-end DNN architecture is constructed by unrolling the alternating minimization solver. Such a novel DNN is built using several layers of unrolled Gauss-Newton iterations followed by proximal steps to account for the regularization term. Interestingly, upon utilizing a GNN-based prior, the power network topology can be exploited in PSSE.  

\subsection{Regularized PSSE with Deep Priors}
In practice, recovering $\bm{v}$ from $\bm{z}$ can be ill-posed, for instance when $\bm{J}_i$ is a rectangular matrix. Building on the data-driven deep priors in image denoising \cite{learningprior2013mri,schlemper2017deep,modl2018tmi}, we advocate regularizing any PSSE loss (here, the NLS in \eqref{eq:exploss}) with a trainable prior information, as 
\begin{equation}\label{eq:denoiser}
\min_{\bm{v} \in \mathbb{R}^{2N}}~\|\bm z-\bm{h}(\bm{v})\|^{2}+\lambda \, \| \bm v - \mathcal D(\bm v) \|^2
\end{equation}
where $\lambda \ge 0$ is a tuning hyper-parameter, while the regularizer promotes states $\bm v$ residing close to ${\mathcal D} (\bm v)$. The latter could be a nonlinear $\hat{\bm v}$ estimator (obtained possibly offline) based on training data. 
To encompass a large family of priors, we advocate a DNN-based estimator $\mathcal D_{\pmb \theta}(\bm v)$ with weights $\pmb \theta$ that can be learned from historical (training) data. Taking a Bayesian view, the DNN $\mathcal{D}_{\pmb \theta}(\cdot)$ can ideally output the posterior mean for a given input.  

Although this regularizer can deal with ill conditioning, the PSSE objective in \eqref{eq:denoiser} remains nonconvex. In addition, the nested structure of $\mathcal D_{\bm \theta}(\cdot)$ presents further challenges. Similar to the Gauss-Newton method for NLS in \eqref{eq:exploss}, we will cope with this challenge using an alternating minimization algorithm to iteratively approximate the solution of \eqref{eq:denoiser}. Starting with some initial guess $\bm{v}_0$, each iteration $i$ uses a linearized data consistency term to obtain the next iterate $\bm{v}_{i+1}$; that is,
\begin{align*}
%\label{eq:gaussdenoiser}
\bm v_{i+1}&= \arg \min_{\bm v} \| \bm{z} - \bm h(\bm v_i)\! -  \bm J_i(\bm v - \bm v_i)\|^2  \!+   \lambda  \| \bm v - \mathcal D_{\pmb \theta} ( \bm v_{i}) \|^2\nonumber\\
&=\bm A_i \bm{z}  +\bm B_i \bm u_{i} + \bm b_i  %\label{eq:dcupdt} 
\end{align*}
where we define 
\begin{subequations}\label{eq:paracacul}
	\begin{align}
	\bm A_i &:=(\bm J^{\top}_i \bm J_i + \lambda \bm I)^{-1} \bm J^{\top}_i  \notag\\
	\bm B_i &:= \lambda (\bm J^{\top}_i \bm J_i + \lambda \bm I)^{-1}  \notag \\
	\bm b_{i} &:=(\bm J^{\top}_i \bm J_i + \lambda \bm I)^{-1} \bm J^{\top}_i  ( \bm J_i \bm v_i- \bm h(\bm v_i) )  \notag.
	\end{align}
\end{subequations}
The solution of \eqref{eq:denoiser} can thus be approached by alternating between the ensuing two steps 
\begin{subequations}\label{eq:update}
	\begin{align}
	\bm u_i &= \mathcal D_{\pmb \theta}(\bm v_{i}) \label{eq:denoisupdt} \\
	\bm v_{i+1} &= \bm A_i \bm{z}  +\bm B_i \bm u_{i} + \bm b_i  \label{eq:dcupdt} .
	\end{align}
\end{subequations}

Specifically, with initialization $\bm v_0 = \bm 0$ and input $\bm{z}$, the first iteration yields $\bm v_1 = \bm A_0 \bm z + \bm B_0 \bm u_0 + \bm b_0$.  Upon passing  $\bm{v}_1$ through the DNN $\mathcal{D}_{\bm{\theta}}(\cdot)$, the output $\bm{u}_1$ at the first iteration, which is also the input to the second iteration, is given by $\bm u_1 = \mathcal D_{\pmb \theta}(\bm v_{1})$ [cf. 
\eqref{eq:denoisupdt}]. In principle, state estimates can be obtained by repeating these alternating iterations whenever a new measurement $\bm{z}$ becomes available. However, at every iteration $i$, the Jacobian matrix $\bm{J}_i$ must be evaluated, followed by matrix inversions to form $\bm{A}_i$, $\bm{B}_i$, and $\bm b_i$. The associated computational burden could be thus prohibitive for real-time monitoring tasks of  large-scale power systems.  

For fast implementation, we pursue an end-to-end learning approach that trains a DNN constructed by unrolling iterations of this alternating minimizer to approximate directly the mapping from measurements $\bm{z}$ to states $\bm{v}$; see Fig. \ref{fig:unrollGNN} for an illustration of the resulting GNU-NN architecture. Recall that in order to derive the alternating minimizer, the DNN prior $\mathcal{D}_{\pmb \theta}(\cdot)$ in \eqref{eq:denoisupdt} was assumed pre-trained, with weights $\pmb \theta$ fixed in advance. In our GNU-NN however, we consider all the coefficients $\{\bm {A}_{i}\}_{i=0}^I $, $\{\bm {B}_{i}\}_{i=0}^I $, $\{\bm {b}_{i}\}_{i=0}^I $, as well as the DNN weights $\{{\bm \theta}_i\}_{i=0}^{I}$ to be learnable from data. 
 
This end-to-end GNU-NN can be trained using backpropagation based on historical or simulated measurements $\{\bm z^t\}_{t=1}^{T}$ and corresponding ground-truth states $\{\bm v^{\ast t}\}_{t=1}^{T}$.
%-state training pairs $\{(\bm z^t, \bm v^{\ast t})\}_{t=1}^{T}$. %\cite{liang2019tsgpsse}.
Entailing only several matrix-vector multiplications, our GNU-NN achieves competitive PSSE performance compared with other iterative solvers such as the Gauss-Newton method. Further, relative to the existing data-driven NN approaches, our GNU-NN can avoid  vanishing and exploding gradients. This is possible thanks to direct (a.k.a skipping) connections from the input layer to intermediate and output layers.

Interestingly, by carefully choosing the specific model for $\mathcal D_{\pmb \theta}(\cdot)$, desirable properties such as scalability and high estimation accuracy can be also effected. For instance, if we use feed forward NNs as $\mathcal D_{\pmb \theta}(\cdot)$, it is possible to obtain a scalable solution for large power networks. However, feed forward NN can only leverage the grid topology indirectly through simulated MATPOWER data. This prompts us to focus on GNNs, which can explicitly capture the topology and the physics of the power network. The resultant Gauss-Newton unrolled with GNN priors (GNU-GNN) is elaborated next.
	
	%%%%%%%%%%%%%%%%%%%%%%%%%%%%%%%%%%%%%%%%%%%%%%%%%%%%%%%%%%%%%%%%%%%%%%%
\subsection{Graph Neural Network Deep Prior}\label{sec:GNNs}
To allow for richly expressive state estimators to serve in our regularization term, we model $\mathcal D_{\pmb \theta}(\cdot)$ through GNNs, that are a prudent choice for networked data. 
GNNs have recently demonstrated remarkable performance in several tasks, including classification, recommendation, and robotics \cite{kipf2016semi, vasileios2019edge}. By operating directly over graphs, GNNs can explicitly leverage the power network topology. Hence, they are attractive options for parameterization in application domains where data adhere to a graph structure~\cite{kipf2016semi,vasileios2020tensor}.   

Consider a graph of $N$ nodes with weighted adjacency matrix $\bm W$ capturing node connectivity. Data matrix $\bm X \in \mathbb R^{N \times F}$ with $n$-th row $\bm{x}_n^\top:=[\bm{X}]_{n:}$ representing an $F\times 1$ feature vector of node $n$, is the GNN input. For the PSSE problem at hand, features are real and imaginary parts of the nodal voltage ($F=2$). Upon pre-multiplying the input $\bm X$ by $\bm W$, features are propagated over the network, yielding a diffused version $\check{\bm Y} \in {\mathbb R}^{N\times F}$ that is given by 
\begin{equation}\label{eq:shiftedallbus}
\check{\bm{Y}}=\bm{W X}.
\end{equation} 
%where %the new graph signal $\bm Y \in {\mathbb R}^{N\times F}$ is a diffused measurement, and 
%$\bm W$ encodes the topological information of the underlying graph. 

\begin{remark}
	To model feature propagation, a common option is to rely on the adjacency matrix or any other matrix that preserves the structure of the power network (i.e.
	$\bm W_{nn'}=0$ if $(n,n') \notin \mathcal E$). Examples include the graph Laplacian, the random walk Laplacian, and their normalized versions.
\end{remark}

Basically, the shift operation in \eqref{eq:shiftedallbus} 
linearly combines the $f$-th features of all neighbors to obtain its propagated feature. Specifically for bus $n$, the shifted feature $[\check{\bm Y}]_{nf}$ is  
\begin{equation}\label{eq:onehopshift}
[\check{\bm{Y}}]_{nf} = \sum_{i=1}^{N}[\bm{W}]_{ni}[\bm{X}]_{if}=\sum_{i \in \mathcal{N}_{n}} w_{ni} x_{i}^{f} 
\end{equation}
where ${\mathcal{N}}_{n}=\{i \in \mathcal{N}:(i, n) \in \mathcal{E}\}$ denotes the set of neighboring buses for bus $n$. Clearly, this interpretation generates a diffused copy or shift of $\bm{X}$ over the graph.

The `graph convolution' operation in GNNs exploits topology information to linearly combine features, namely
\begin{equation}
\label{eq:gc}
[\bm{Y}]_{nd}:=[\mathcal{H} \star \bm{X}; \bm{W}]_{nd}:=\sum_{k=0}^{K-1}[\bm{W}^{k} \bm{X}]_{n:} [\bm{H}_k]_{:d} 
\end{equation}
where $\mathcal{H}:=[\bm{H}_0~ \cdots~\bm{H}_{K-1}]$ with ${\bm H}_{k} \in \mathbb{R}^{F\times D}$ concatenating all filter coefficients;  $\bm Y \in \mathbb R^{N\times D}$ is the intermediate (hidden) matrix with $D$ features per bus; and,   
$\bm{W}^{k} \bm{X}$ linearly combines features of buses within the $k$-hop neighborhood by recursively applying the shift operator $\bm W$.

\begin{figure}
	\centering
	\includegraphics[width =0.45 \textwidth]{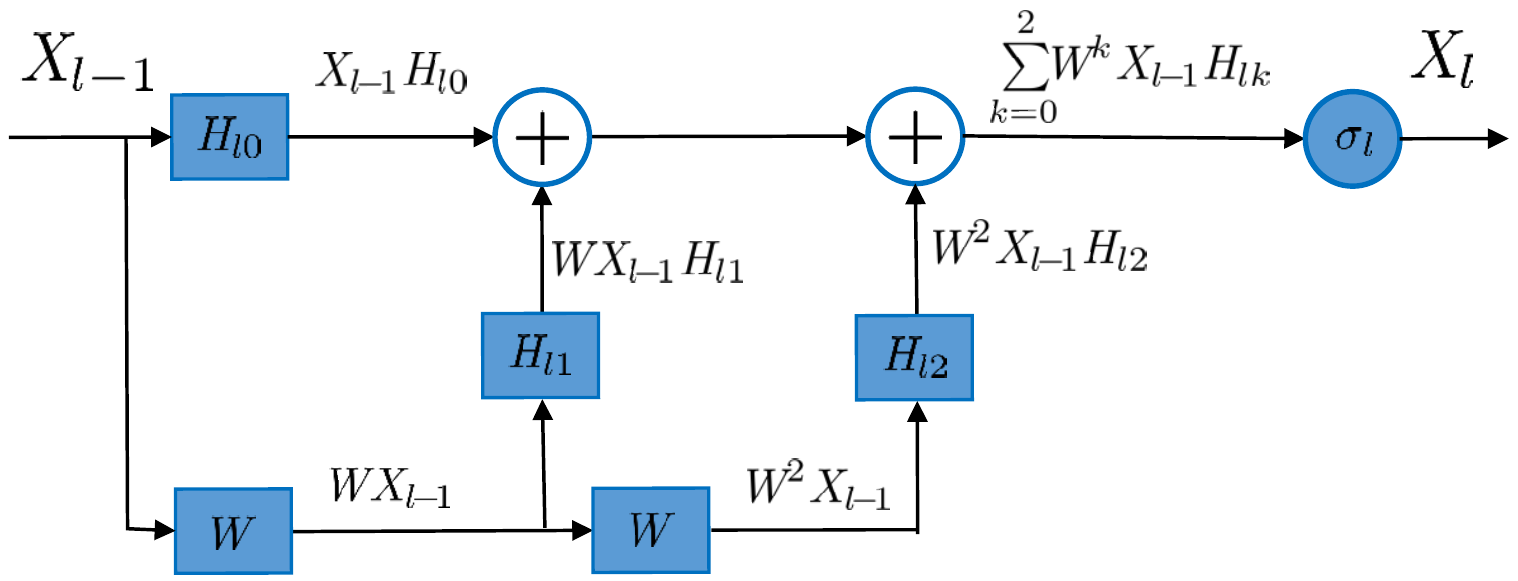}
	\caption{The signal diffuses from layer $l-1$ to $l$ with $K= 3$.}
	\label{fig:GNNflow}
\end{figure}

To obtain a GNN with $L$ hidden layers, let $\bm{X}_{l-1}$ denote the output of the $(l-1)$-st layer, which is also the $l$-th layer input for $l=1, \ldots, L$, and $\bm{X}_0 = \bm{X}$ is the input matrix. The hidden $\bm{Y}_{l} \in \mathbb{R}^{N\times D_{l}}$ with $D_l$ features is obtained by applying the graph convolution operation \eqref{eq:gc} at layer $l$, that is
\begin{equation}
\label{eq:gclayer} 
[\bm{Y}_l]_{nd} =\sum_{k=0}^{K_l-1}[\bm{W}^{k} \bm{X}_{l-1}]_{n:} [\bm{H}_{lk}]_{:g} 
\end{equation}
where $\bm{H}_{lk} \in \mathbb{R}^{F_{l-1}\times F_{l}}$ are the graph convolution coefficients for $k=0, \ldots, K_l-1$. The output $\bm X_l$ at layer $l$ is found by applying a graph convolution followed by a point-wise nonlinear operation $\sigma_{l}(\cdot)$, such as the rectified linear unit (ReLu) $\sigma_{l}(t):=\max\{0,\,t\} $ for $t\in\mathbb{R}$; see Fig. \ref{fig:GNNflow} for a depiction. Rewriting \eqref{eq:gclayer} in a compact form, we arrive at  
\begin{equation}\label{llayershift}
\bm{X}_{l} =\sigma_{l}(\bm Y_{l})=\sigma_{l}\!\left(\sum_{k=0}^{K_l-1} \bm{W}^{k} \bm{X}_{l-1} \bm{H}_{l k}\right).
\end{equation}
The GNN-based PSSE provides a nonlinear functional operator $\bm{X}_L=\bm { \Phi}(\bm{X}_{0} ; \bm{\Theta}, \bm{W})$ that maps the GNN input $\bm{X}_{0}$ to voltage estimates by taking into account the graph structure through $\bm W$, through
\begin{align}
&\bm { {\Phi}}(\bm{X}_{0} ; \bm{\Theta}, \bm{W}) = \label{eq:GNNbasepsse} \\ 
&\sigma_{L}\!\left(\sum_{k=0}^{K_{L }-1} \bm{W}^{k} \!\left(\ldots \!\left(\sigma_{1}\!\left(\sum_{k=0}^{K_1-1} \bm{W}^{k} \bm{X}_{0} \bm{H}_{1 k}\right) \ldots\right)\right)\bm{H}_{L k}\right)\nonumber
\end{align}
where the parameter set $\bm \Theta$ contains all the filter weights; that is, $\bm{\Theta} := \{\bm{H}_{lk}, \forall l, k\}$, and also recall that $\bm X_0 = \bm X$.

\begin{remark}
With $L$ hidden layers, $F_l$ features and $K_l$ filters per layer, the total number of parameters to be learned is $|\bm{\Theta}| =\sum_{l=1}^{L} K_{l} \times F_{l} \times F_{l-1}$. 
\end{remark}

To accommodate the GNN implementation over the proposed unrolled architecture, at the $i$-th iteration,  we reshape the states $\bm{v}_{i} \in \mathbb{R}^{2N}$ to form the $N \times 2$ GNN input matrix $\bm X_0^{i} \in {\mathbb R}^{N \times 2}$. Next, we vectorize the GNN output $\bm{X}_L^i \in \mathbb{R}^{N\times 2}$ to obtain the vector $\bm u_i \in \mathbb{R}^{2N}$ (cf. \eqref{eq:denoisupdt}). For notational brevity, we concatenate all trainable parameters of the GNU-GNN in vector $\bm{\omega}:=[\{\bm{\Theta}_i\}_{i=0}^{I}, \{{\bm A}_i\}_{i=0}^{I}, \{\bm{B}_i\}_{i=0}^{I}, \{\bm{b}^1_i\}_{i=0}^{I}]$, and let $\bm{\pi}(\bm{z};\bm{\omega})$ denote the end-to-end GNU-GNN parametric model, which for given measurements $\bm{z}$ predicts the voltages across all buses, meaning $\hat{\bm{v}}=\bm{\pi}(\bm{z};\bm{\omega})$. The GNU-GNN weights $\bm \omega$ can be updated using backpropagation, after specifying a certain loss $\ell(\bm{v}^{\ast}, \bm{v}_{I+1})$ measuring how well the estimated voltages $\bm{v}_{I+1}$ by the GNU-GNN matches the ground-truth ones $\bm{v}^{\ast}$. The proposed method is summarized in Alg. \ref{Alg_a}.

\begin{algorithm}[t!]
	\caption{PSSE Solver with GNN Priors.}
	\label{Alg_a}
	\hspace*{0.02in} {\bf Training phase:}
	\begin{algorithmic}[1]
		\State \textbf{Input:} Training samples $\{(\bm z^t, {\bm v}^{\ast t})\}_{t=1}^{T}$
		\State \textbf{Initialize:} 
		\newline $\bm{\omega}^1 :=[\{\bm{\Theta}^1_i\}_{i=0}^{I}, \{{\bm A}^1_i\}_{i=0}^{I}, \{\bm{B}^1_i\}_{i=0}^{I}, \{\bm{b}^1_i\}_{i=0}^{I}]$, $\bm v_0=0$.
		%		$\{\bm A^0_i\}^I_{i=1}$, $\{\bm B^0_i\}^I_{i=1}$, $\{\bm u^t_0\}^T_{t=1}$, and $\{\bm \Theta^0_i\}^I_{i=1}$.
		\For{$t =1,2,\ldots, T$ } 
		\State Feed $ \bm z^t$  and $\bm v_{0}$ as input into GNU-GNN.
		\For{$i = 0, 1, \ldots, I$}\footnotemark
		\State Reshape $\bm v_i \in \mathbf R^{2N}$ to get $\bm X_0^{i} \in {\mathbb R}^{N \times 2}$.
		\State Feed $\bm X_0^{i}$ into GNN.
		\State Vectorize the GNN output $\bm{X}_L^i \in \mathbb{R}^{N\times 2}$ to get $\bm u_i$.
		\State Obtain $\bm v_{i+1} \in \mathbf R^{2N}$ using \eqref{eq:dcupdt}.
		\EndFor
		\State Obtain $\bm v^t_{I+1}$ using \eqref{eq:dcupdt}.
		\State Minimize the loss $\ell(\bm{v}^{\ast t}, \bm{v}^t_{I+1})$ and update $\bm{\omega}^{t}$.
		\EndFor
		\State \textbf{Output:} $\bm{\omega}^{T}$
	\end{algorithmic}
	\hspace*{0.02in} {\bf Inference phase:} 
	\begin{algorithmic}[1]
		\For {$t = T+1, \ldots, T'$}
		\State Feed real-time $\bm{z}^t$ to the trained GNU-GNN.
		\State Obtain the estimated voltage $\bm v^{t} $.
		\EndFor
	\end{algorithmic}
	
\end{algorithm}

%%%%%%%%%%%%%%%%%%%%%%%%%%%%%%%%%%%%%%%%%%%%%%%%%%%%%%%%%%%%%%%%%%%%%%%
	\section{Robust PSSE Solver}\label{sec:adve}
In real-time inference, our proposed GNU-GNN that has been trained using past data, outputs an estimate of the state $\bm v^t$ per time slot $t$ based on the observed measurements $\bm z^t$.
However, due to impulsive communication noise and possibly cyberattacks,  our proposed GNU-GNN in Section \ref{sec:unro} can yield grossly biased estimation results. To obtain estimators robust to bad data, classical formulations including H\"uber estimation, H\"uber M-estimation, and Schweppe-H\"uber generalized M-estimation, rely on the premise that measurements obey $\epsilon$-contaminated probability models; see e.g., \cite{mili1994robust,wang2019overview}. Instead, the present paper postulates that measured and ground-truth voltages are drawn from some nominal yet unknown distribution $P_0$ supported on $\mathcal{S} = \mathcal{Z}\times \mathcal{V}$, that is $(\bm{z}, \bm{v}^{\ast}) \sim P_0$. 
%Under this distributional assumption, 
Therefore, to obtain the end-to-end GNU-GNN parametric model $\bm
{\pi}(\bm{z};\bm{\omega})$, the trainable parameters $\bm{\omega}$ are optimized by solving $
{\min}_{\,\bm{\omega}}~\mathbb{E}_{P_{0}}\big[\ell(\bm{\pi}(\bm{z};\bm{\omega}),\bm{v}^{\ast })\big]$ \cite{miller2019adversarial}. In practice, $P_0$ is unknown but i.i.d. training samples $\{ (\bm z^t, \bm v^{\ast t})\}_{t=1}^{T} \sim P_0$ are available. In this context, our PSSE amounts to solving for the minimizer of the empirical loss as
\begin{equation} \label{eq:emploss}
\min_{\bm{\omega}}\; \bar{\mathbb{E}}_{\widehat{P}^{(T)}_0} [\ell(\bm{\pi}(\bm{z}^t;\bm{\omega}),\bm{v}^{\ast t})] =  \frac{1}{T} \sum_{t=1}^T \ell(\bm{\pi}(\bm{z}^t;\bm{\omega}),\bm{v}^{\ast t}).
\end{equation} 
\footnotetext{For brevity the superscript $t$ is removed from inner iteration $i$.} 
To cope with uncertain and adversarial environments, the solution of \eqref{eq:emploss} can be robustified by optimizing over a set $\mathcal{P}$ of probability distributions centered around    $\widehat{P}^{(T)}_0$, and minimizing the \textit{worst-case} expected loss with respect to the choice of any distribution $P \in \mathcal{P}$. Concretely, this can be formulated as the following \textit{distributionally robust} optimization
\begin{equation}\label{eq:worstcase}
\min_{\bm{\omega}} \sup _{P \in \mathcal{P}}\; \mathbb{E}_{P}[\ell(\bm{\pi}(\bm{z};\bm{\omega}),\bm{v}^\ast)].
\end{equation}
Compared with \eqref{eq:emploss}, the worst-case formulation in \eqref{eq:worstcase} ensures a reasonable performance across a continuum of distributions in $\mathcal{P}$. A broad range of ambiguity sets $\mathcal{P}$ could be considered here. Featuring a strong duality enabled by the optimal transport theory \cite{vinali08opttrans}, such distributionally robust optimization approaches have gained popularity in robustifying machine learning models \cite{bandi2014robust}. Indeed, this tractability is the key impetus for this section.

%including momentum \cite{}, KL divergence \cite{}, statistical test \cite{}, and currently popular Wasserstein distance-based sets \cite{}; see e.g., \cite{} for a thorough discussion. 

%The optimal transport theory was first studied by Monge \cite{monge1781memoire}. In the Monge problem, piles of sand and some holes with the same total volume as sands are geographically distributed over an area. 
%The goal of optimal transport is to find the optimal moves to transfer the entire piles to the holes with the minimum transportation cost. Clearly,  the optimal strategy depends on the cost function, which is a function of distance between piles and holes.

To formalize, consider probability density functions $P$ and $Q$ defined over support $\mathcal{S}$, and let $\Pi(P,Q)$ be the set of all joint probability distributions with marginals $P$ and $Q$. Also let $c: \mathcal{Z} \times \mathcal{Z}  \rightarrow [0, \infty)$ be some cost function representing the cost of transporting a unit of mass from $(\bm{z}, \bm{v}^\ast)$ in $P$ to another element $(\bm{z}', \bm{v}^\ast)$ in $Q$ (here we assume that attacker can compromise the measurements $\bm{z}$ but not the actual system state $\bm{v}^\ast$). The so-called optimal transport between two distributions $P$ and $Q$ is given by \cite[Page 111]{vinali08opttrans}
\begin{align}
W_c(P,Q) := \; \underset{\pi \in \Pi}{\inf} \, \mathbb{E}_\pi \big[ c(\bm{z},\bm{z}')\big]. 
\label{eq:wassdist}
\end{align}
Intuitively, $W_c(P,Q)$ denotes the minimum cost associated with transporting all the mass from distribution $P$ to $Q$. Under mild conditions over the cost function and distributions, $W_c$ gives the well-known Wasserstein distance between $P$ and $Q$; see e.g., \cite{sinha2017certify}.
%\begin{remark}
%	If $c(\cdot)$ satisfies the axioms of distance, then $W_c$ defines a distance on the space of probability measures. For instance, if $P$ and $Q$ are defined over a Polish space equipped with metric $d$, then choosing $c(\bm z, \bm z') = d^p(\bm z, \bm z')$ for some $p\in [1, \infty)$ asserts that $W_c^{1/p}(P,Q)$ is the well-known Wasserstein distance of order $p$ between probability measures $P$ and $Q$ \cite[Definition 6.1]{vinali08opttrans}.  
%\end{remark}

Having introduced the distance $W_c$, let us define an uncertainty set for the given empirical distribution $\widehat{P}^{(T)}_0$, as $\mathcal{P}:= \{P| W_c(P,\widehat{P}^{(T)}_0) \le \rho \}$ that includes all probability distributions having at most $\rho$-distance from $P_0^{(T)}$. Incorporating $\mathcal{P}$ into \eqref{eq:worstcase} yields the following optimization for distributionally robust GNU-GNN estimation
\begin{subequations}\label{eq:robform}
\begin{align}\label{eq:robformbobj}
\min_{\bm{\omega}} \, \sup _{P}\,&~ \mathbb{E}_{P}[\ell(\bm{\pi}(\bm{z};\bm{\omega}),\bm{v}^\ast)]\\
%\end{align}
%\begin{equation}
	\label{eq:robformbcons}
\qquad {\rm s.t.} &~ W_c(P,\widehat{P}^{(T)}_0) \le \rho. 
\end{align}
\end{subequations}
Observe that the inner functional optimization in \eqref{eq:robformbobj} runs over all  probability distributions $P$ characterized by \eqref{eq:robformbcons}. Evidently, optimizing directly over the infinite-dimension distribution functions is intractable. Fortunately, for continuous loss as well as transportation cost functions, the inner maximization satisfies strong duality condition; that is, the optimal objective value of the inner maximization is equal to its dual optimal objective value. In addition, the dual problem involves optimization over only a one-dimension variable, that can be carried out efficiently. These two observations prompt us to solve \eqref{eq:robform} in the dual domain. 
To formally obtain this tractable surrogate, we call for a result from \cite{blanchet2019quantifying}.

\begin{proposition}
\label{prop:strongdual}
Let the loss $\ell: \bm{\omega} \times \mathcal{Z} \times \mathcal{V} \rightarrow [0,\infty)$, and transportation cost $c:\mathcal{Z} \times \mathcal{Z} \rightarrow [0, \infty)$ be continuous functions. Then, for any given $\widehat{P}^{(T)}_0$, and $\rho > 0$, it holds \begin{align} \label{eq:strongduality}
\sup_{P\in\mathcal{P}} & \,\mathbb{E}_{P}\!\left[\ell(\bm{\pi}(\bm{z};\bm{\omega}), \bm{v}^\ast)
\right] =  \\ 
& \inf_{\gamma \ge 0}  \big\{  \bar{\mathbb{E}}_{{(\bm{z}, \bm{v}^\ast)\sim\widehat{P}^{(T)}_0}} \big[\sup_{\bm \zeta \in {\mathcal Z}}  \ell(\bm{\pi}(\bm{\zeta};\bm{\omega}),\bm{v}^\ast) \!+\! \gamma (\rho - c(\bm z, \bm \zeta)) \big] \big\} 
\nonumber
\end{align}
where $\mathcal{P}:= \left\{P| W_c(P,\widehat{P}^{(T)}_0) \le \rho \right\}$.
\end{proposition}

\begin{remark}
	Thanks to the strong duality, the right-hand side in \eqref{eq:strongduality} simply is a univariate dual reformulation of the primal problem given on the left-hand side. In contrast with the primal formulation, the expectation in the dual domain is taken only over the empirical distribution $\widehat{P}^{(T)}_0$ rather than over any $P \in \mathcal{P}$. Furthermore, since this reformulation circumvents the need for finding the optimal coupling $\pi \in \Pi$ to define $\mathcal{P}$, and  characterizing the primal objective for all $P \in \mathcal{P}$, it is practically appealing and convenient.
\end{remark}

Capitalizing on Proposition \ref{prop:strongdual}, we can replace the inner maximization with its dual reformulation to arrive at the following distributionally robust PSSE optimization
\begin{align}
\min_{\bm \omega} \inf_{\gamma \ge 0}   {\bar{\mathbb{E}}}_{(\bm{z}, \bm{v}^\ast)\sim\widehat{P}^{(T)}_0} \Big[\sup_{\bm \zeta \in {\mathcal Z}} \ell(\bm{\pi}(\bm{\zeta};\bm{\omega}),\bm{v}^\ast) + \gamma (\rho - c(\bm z, \bm \zeta)) \Big]\!.
\label{eq:robustdual}
\end{align}
   
%Before proceeding to solve this learning problem in the next section, two remarks are in order. 

\begin{remark}
\label{rm:minmax}
Although the robust surrogate in \eqref{eq:robustdual} looks similar to minimax (saddle-point) optimization problems, it requires the supremum to be solved separately per observed measurements $\bm{z}$, that cannot readily be handled by existing minimax optimization solvers. 
\end{remark} 
%As remark \ref{rm:minmax} suggests, the problem of interest in \eqref{eq:robustdual} cannot be handle through existing optimization techniques. 

Finding the optimal solution $(\bm{\omega}^\ast,\gamma^\ast)$ of 
 \eqref{eq:robustdual} is in general challenging. A common approach to bypassing this hurdle is to approximate the optimal $\bm{\omega}^\ast$ by solving  \eqref{eq:robustdual} with a preselected and fixed $\gamma>0$  \cite{sinha2017certify}. Indeed, it has been shown in \cite{sinha2017certify} that for any strongly convex transportation cost function, such as $c(\bm{z}, \bm{z'}) :=\| \bm{z} - \bm{z}'\|^2_p$ for any $p \ge 1$, a sufficiently large $\gamma>0$ ensures that the inner maximization is strongly convex, hence efficiently solvable.  
% 
% 
%This is because the objective function resultant from inner maximization may become non differentiable and nonconvex. The remedy is to fix the penalty $\gamma \ge 0$ to appropriate value so that the inner maximization admits a unique solution. Consequently, iterative solvers for  \eqref{eq:robustdual} can be proposed. These algorithms essentially iterate between minimizing over $\bm{\omega}$ and maximizing over $\bm{\zeta}$, and provably converge to a stationary point \cite{sinha2017certify}. This is an immediate result of Danskin's theorem for minimax optimization problems; see \cite{danskin1966theory} for more details. 
%%\gang{It has been shown that with large enough $\gamma>0$, one can approximate the optimal solution $\bm{\omega}^\ast$ of \eqref{eq:robustdual} iteratively by alternating between minimizing over $\bm{\omega}$ and maximizing over $\bm{\zeta}$ \cite{sinha2017certify}. 
%%Unfortunately, such an approach leads to suboptimal performance as it is difficult to set the optimal $\gamma^\ast$. Certainly, this could be done through cross-validation, which however will incur significant computational burden. 
%Prompted by this result, our approach is to employ a strongly convex function as the transportation cost, such as $c(\bm{z}, \bm{z'}) :=\| \bm{z} - \bm{z}'\|^2_p$ for any $p \ge 1$. Upon appropriately choosing $\gamma$, the inner maximization has a unique solution, therefore the Danskin's theorem can be invoked. As a consequence, we employ an alternating solver for \eqref{eq:robustdual}.   
Note that having a fixed $\gamma$ is tantamount to tuning $\rho$, which in turn \emph{controls} the level of infused \emph{robustness}. Fixing some large enough $\gamma > 0$ in \eqref{eq:robustdual}, our robustified GNU-GNN model can thus be obtained by solving  
\begin{equation}
\min_{\bm{\omega}}~\bar{\mathbb E}_{{(\bm{z}, \bm{v}^\ast)\sim\widehat{P}^{(T)}_0}}\Big[ \sup_{\bm{\zeta}\in\mathcal{Z}} \psi(\bm{\bm{\omega}}, \bm{\zeta}; \bm {z},\bm{v}^\ast) \Big]
\label{eq:objective}
\end{equation}
where 
\begin{equation} 
\label{eq:psi}
\psi(\bm{\bm{\omega}}, \bm{\zeta}; \bm {z},\bm{v}^\ast):=\ell(\bm{\pi}(\bm{\zeta};\bm{\omega}),\bm{v}^\ast) + \gamma (\rho -  c(\bm{z}, \bm \zeta)).
\end{equation}

Intuitively, \eqref{eq:objective} can be understood as first `adversarially' perturbing the measurements $\bm{z}$ into $\bm{\zeta}^\ast$ 
by maximizing $\psi(\cdot)$, and subsequently seeking a model that minimizes the empirical loss with respect to even such perturbed inputs. In this manner, robustness of the sought model is achieved to future data that may be contaminated by adversaries.

Initialized with some $\bm{\omega}^0$, and given a datum $(\bm{z}^t, \bm{v}^\ast)$, we form $\psi(\cdot)$ (c.f. \eqref{eq:psi}), and implement a single gradient ascent step for the inner maximization as follows 
\begin{align}
\bm{\zeta}^t =  \bm{z}^t  + \eta^t \nabla_{\bm{\zeta}} \psi({\bm \omega}^t, \bm  \zeta; \bm{z}^t, \bm{v}^{\ast t})\big|_{\bm{\zeta}={\bm{z}^t}} \label{eq:sgaupdt}
\end{align}
where $\eta^t>0$ is the stepsize.
Upon evaluating \eqref{eq:sgaupdt}, the perturbed data $\bm{\zeta}^t$ will be taken as input (replacing the `healthy' data $\bm{z}^t$) fed into Algorithm \ref{Alg_a}. Having now the loss $\ell(\bm{\pi}(\bm{\zeta}^t;\bm{\omega}),\bm{v}^{\ast t})$ as solely a function of the GNU-GNN weights $\bm{\omega}$, the current iterate $\bm{\omega}^t$ can be updated again by means of backpropogation.

%%%%%%%%%%%%%%%%%%%%%%%%%%%%%%%%%%%%%%%%%%%%%%%%%%%%%%%%%%%%%%%%%%%%%%% 
\section{Numerical Tests} \label{sec:test}
This section tests the estimation performance as well as robustness of our proposed methods on the IEEE $118$-bus benchmark system.

\subsection{Simulation Setup}	

The simulations were carried out on an NVIDIA Titan X GPU with a $12$GB RAM. For numerical tests, we used real load consumption data from the $2012$ Global Energy Forecasting Competition (GEFC) \cite{datagefc}. Using this dataset, training and testing collections were prepared by solving the AC power flow equations using the MATPOWER toolbox \cite{MATPOWER}. To match the scale of power demands, we normalized the load data, and fed it into MATPOWER to generate $1,000$ pairs of measured and ground-truth voltages, $80\%$ of which were used for training while the remaining $20\%$ were employed for testing. Measurements include all sending-end active power flows, as well as voltage magnitudes, corrupted by additive white Gaussian noise. Standard deviations of the noise added to power flows and voltage magnitudes were set to $0.02$ and $0.01$ \cite{gang2019tsg}, respectively.

A reasonable question to ponder is whether explicitly incorporating the power network topology through a trainable regularizer offers improved performance over competing alternatives. In addition, it is of interest to study how a distributionally robust training method enhances PSSE performance in the presence of bad data and even adversaries. To this aim, three baseline PSSE methods were numerically tested, namely: i) the prox-linear network in~\cite{liang2019tsgpsse}; ii) a $6$-layer vanilla feed-forward (F)NN; and iii) an $8$-layer FNN. The weights of these NNs were trained using the `Adam' optimizer to minimize the H\"uber loss. The learning rate was fixed to $10^{-3}$ throughout $500$ epochs, and the batch size was set to $32$. 

\begin{figure}
	\centering
	\includegraphics[width =0.42 \textwidth]{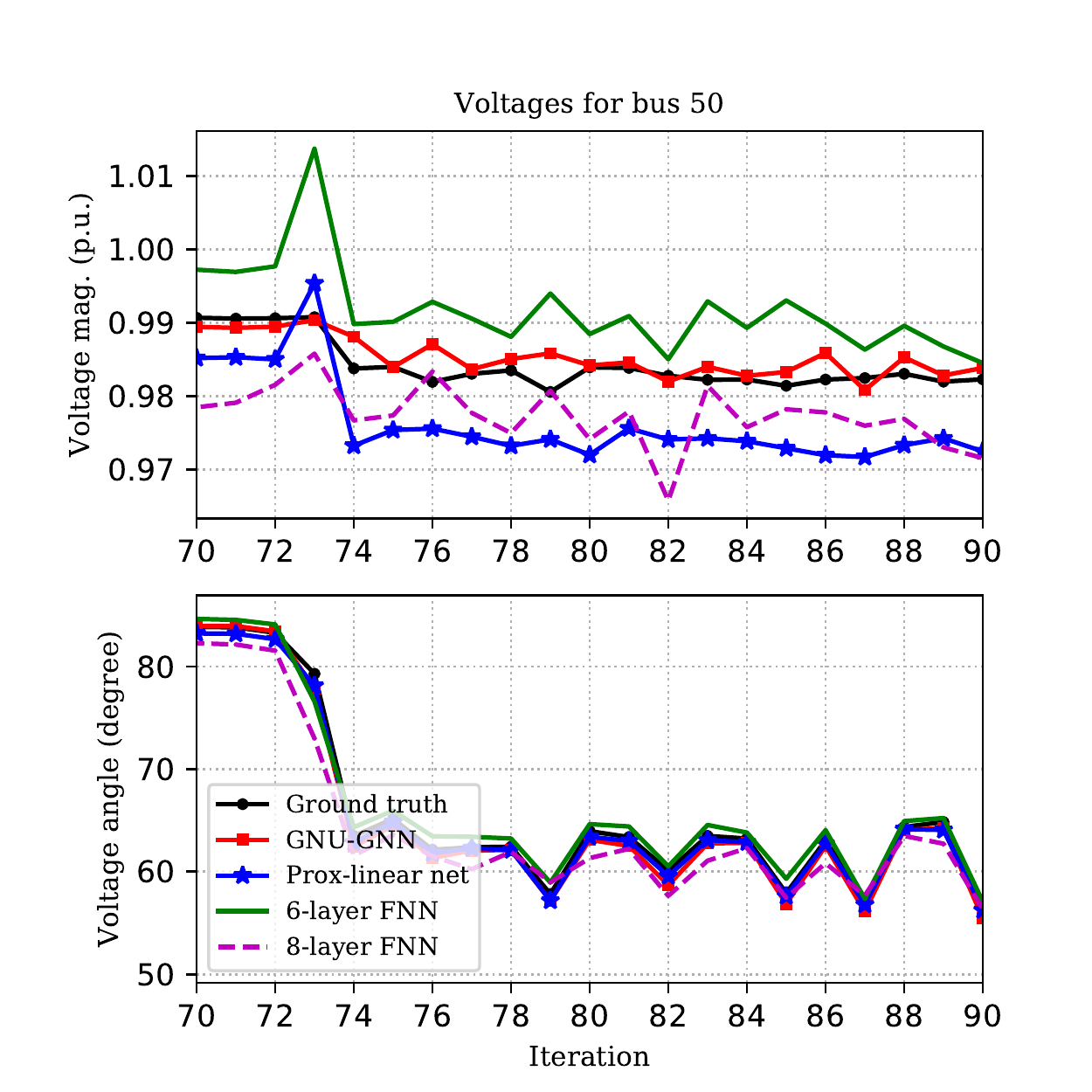}
	\caption{The estimated voltage magnitudes and angles by the four schemes at bus $50$ from slots $70$ to $90$.}
%	\vspace{-.2 in}
	\label{fig:vol_clean_50}
\end{figure}

\begin{figure}
	\centering
	\includegraphics[width =0.42 \textwidth]{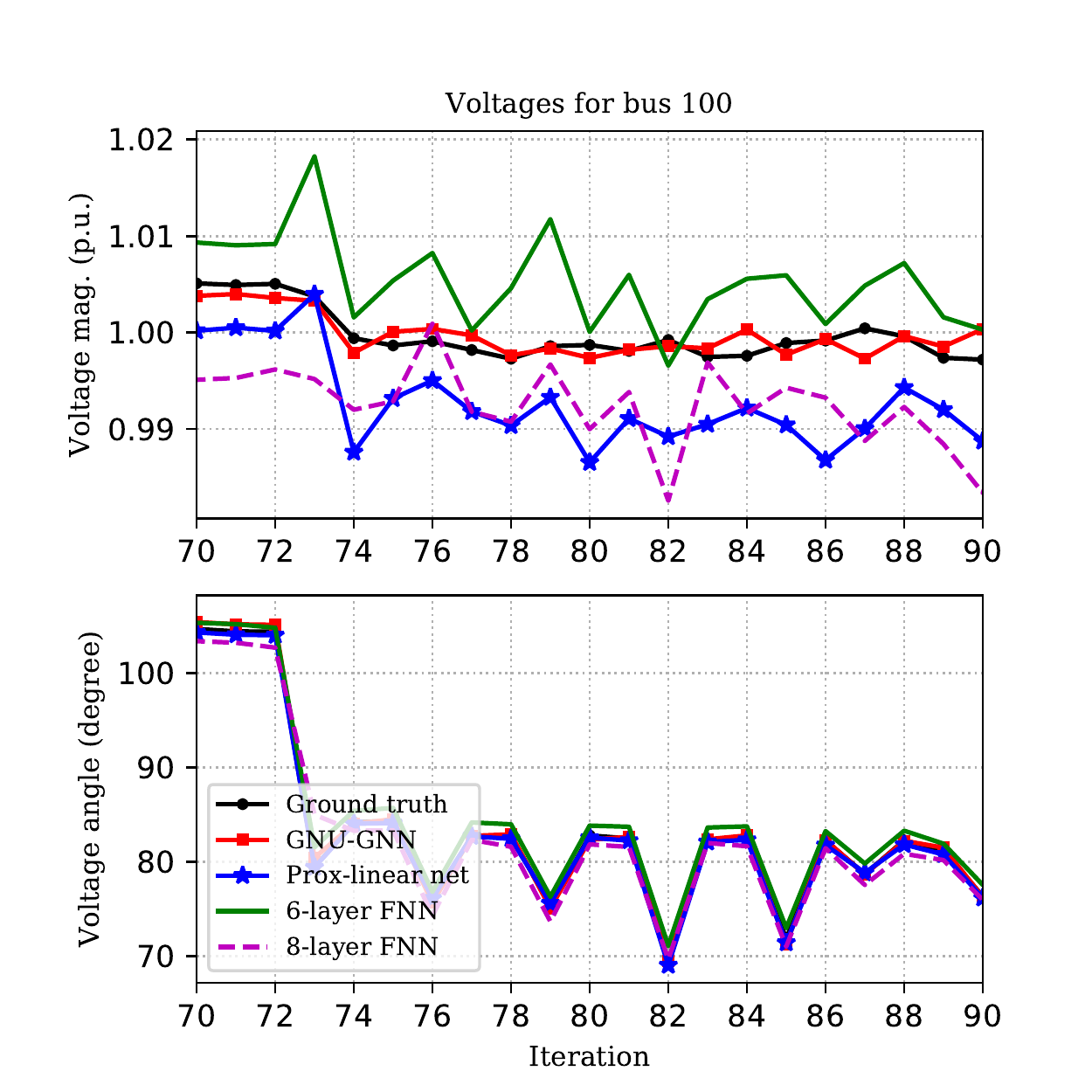}
	\caption{The estimated voltage magnitudes and angles by the four schemes at bus $100$ from slot $70$ to $90$.}
	\label{fig:vol_clean_100}
\end{figure}

\subsection{GNU-GNN and GNU-FNN for regularized PSSE}
In the first experiment, we implemented GNU-GNN by unrolling $I=6$ iterations of the proposed alternating minimizing solver, respectively. A GNN with $K=2$ hops, and $D=8$ hidden units with ReLU activations per unrolled iteration was used for the deep prior of GNU-GNN. The GNU-GNN architecture was designed to have total number of weight parameters roughly the same as that of the prox-linear network. 

\begin{figure}[h!]
	\centering
	\includegraphics[width =0.42 \textwidth]{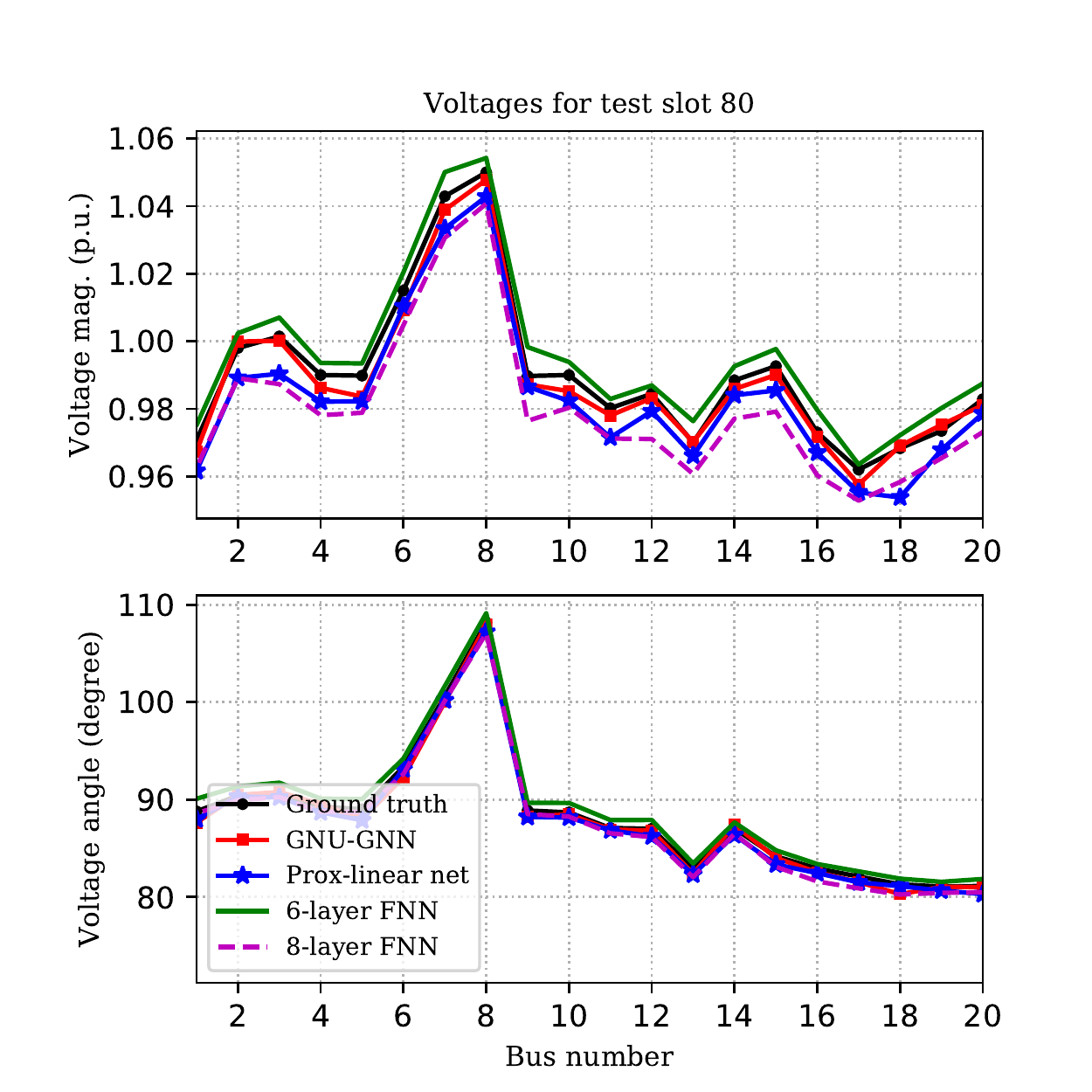}
	\caption{The estimated voltages magnitudes and angles by the four schemes for the first $20$ buses at slot $80$.}
	\label{fig:vol_clean_allbus}
			\vspace{-1em}
\end{figure}

The first set of results depicted in Figs. \ref{fig:vol_clean_50} and \ref{fig:vol_clean_100} show the estimated voltage profiles obtained at buses $50$ and  $100$ from test slots $70$ to $90$, respectively. The ground-truth and estimated voltages for the first $20$ buses on the test slot $80$ are presented in Fig. \ref{fig:vol_clean_allbus}. These plots corroborate the improved performance of the our GNU-GNN relative to the simulated PSSE solvers. 

%\qiu{To further investigate the impact of our trainable regularization approach, we adopted FNN as prior and formed a GNU-FNN architecture. The FNNs all have one hidden layer with ReLU activation functions. The estimated voltage profiles are reported in Figs. \ref{fig:vol_clean_50} and \ref{fig:vol_clean_100}. Interestingly, despite  using  topology information implicitly, the GNU-FNN performance is comparable with the GNU-GNN in which uses grid topology explicitly.}   

\begin{figure}
	\centering
	\includegraphics[width =0.42 \textwidth]{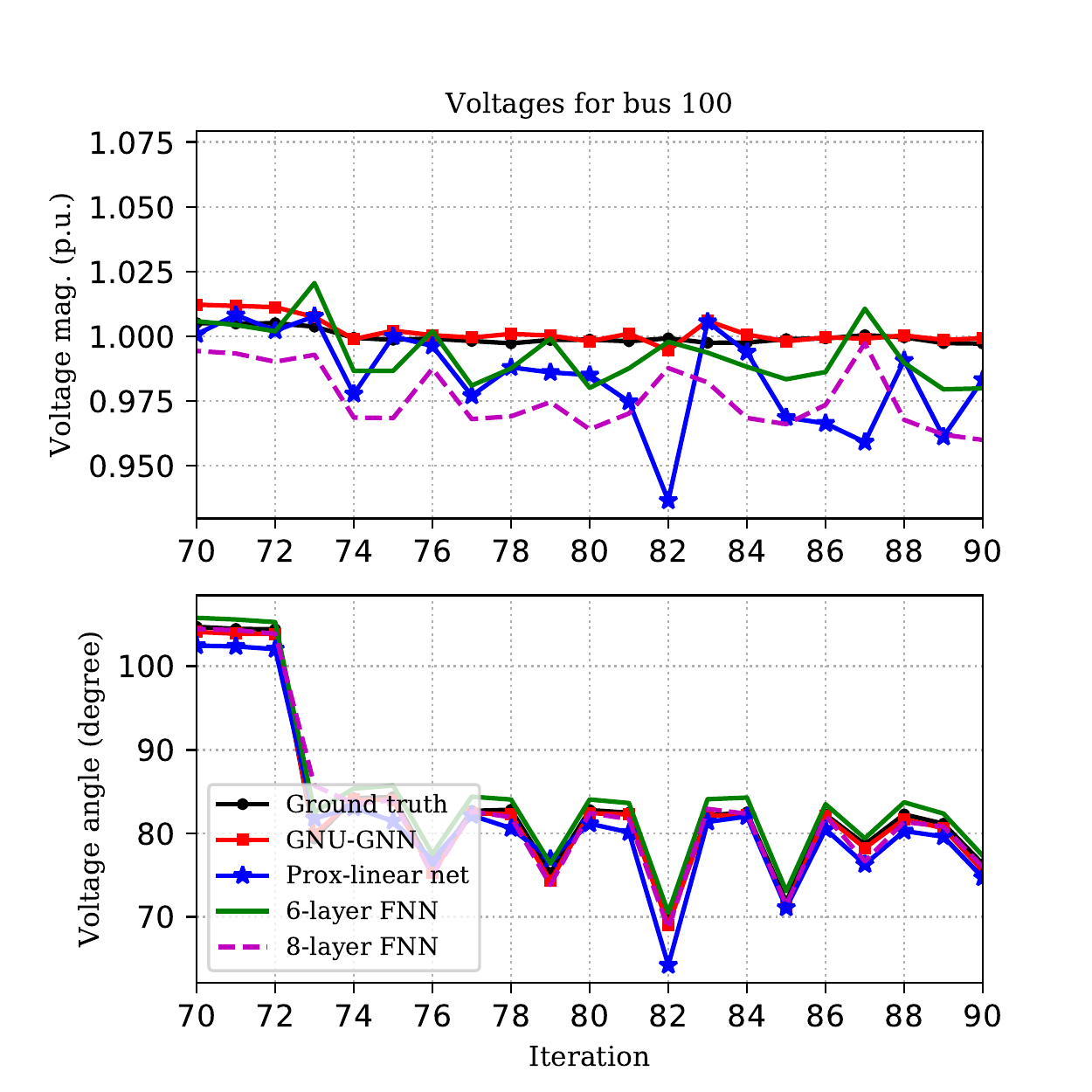}
	\caption{The estimated voltage magnitudes and angles
	by the four schemes under distributional attacks at bus $100$ from slots $70$ to $90$.}
	\label{fig:vol_advdistr_80}
\end{figure}

\subsection{Robust PSSE}
\begin{figure}
	\centering
	\includegraphics[width =0.42 \textwidth]{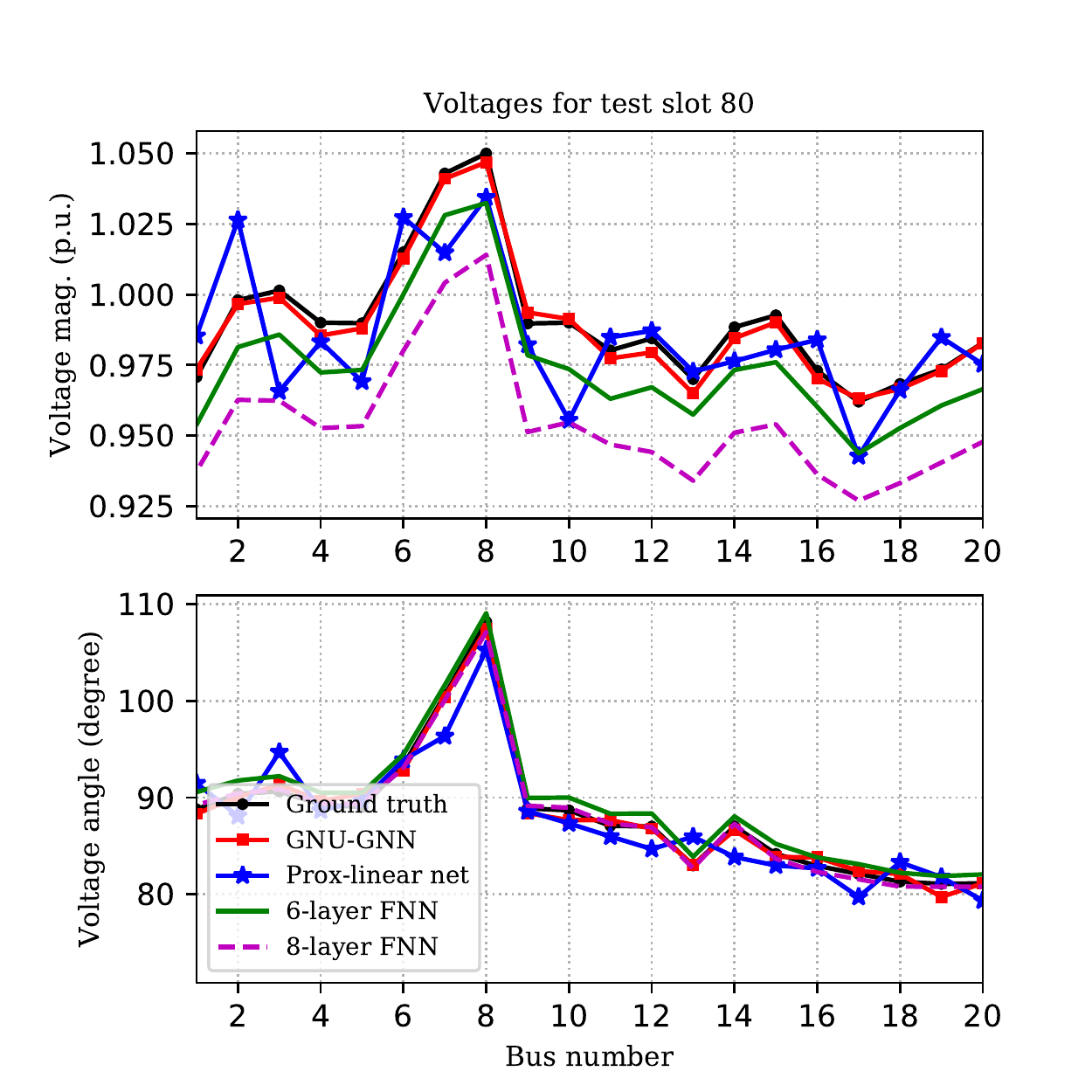}
	\caption{The estimated voltage magnitudes and angles by the four schemes under distributional attacks for the first $20$ buses at slot $80$. }
	\label{fig:vol_advdistr_slot50}
\end{figure}

Despite their remarkable performance in standard PSSE, DNNs may fail to yield reliable and accurate estimates in practice when bad data are present. Evidently, this challenges their application in safety-critical power networks. In the experiment of this subsection we examine the robustness of our GNU-GNN trained with the described adversarial learning method. To this aim, a distributionally robust learning scheme was implemented to manipulate the input of GNU-GNN, prox-linear net, $6$-layer FNN, and $8$-layer FNN models. Specifically, under distributional attacks, an ambiguity set $\mathcal{P}$ comprising distributions centered at the nominal data-generating $P_0$ was postulated. Although the training samples were generated according to $P_0$, testing samples were obtained by drawing samples from a distribution $P \in \mathcal{P}$ that yields the worst empirical loss. To this end we preprocessed test samples using \eqref{eq:sgaupdt} to generate adversarially perturbed samples.  Figs. \ref{fig:vol_advdistr_80} and \ref{fig:vol_advdistr_slot50} demonstrate the estimated voltage profiles under a distributional attack with a fixed $\gamma = 0.13$ (c.f. \eqref{eq:objective} and \eqref{eq:psi}) and $\ell_2$ transportation cost. As the plots showcase, our proposed robust training method enjoys guarantees against distributional uncertainties, especially relative to competing alternatives. 
%For relatively small values of $\gamma$ (large adversarial budgets), our method is a heuristic way to  Here we have chosen $\gamma$ 

%%%%%%%%%%%%%%%%%%%%%%%%%%%%%%%%%%%%%%%%%%%%%%%%%%%%%%%%%%%%%%%%%%%%%%%
\section{Conclusions}\label{sec:conc}
This paper introduced topology-aware DNN-based regularizers to deal with the ill-posed and nonconvex characteristics of standard PSSE approaches. An alternating minimization solver was developed to approach the solution of the regularized PSSE objective function, which is further unrolled to construct a DNN model. For real-time monitoring of large-scale networks, the resulting DNN was trained using historical or simulated measured and ground-truth voltages. A basic building block of our GNU-GNN consists of a Gauss-Newton iteration followed by a proximal step to deal with the regularization term. Numerical tests showcased the competitive performance of our proposed GNU-GNN relative to several existing ones. Further, a distributioally robust training method was presented to endow the GNU-GNN with resilience to bad data that even come from adversarial attacks. 
 
Future directions include investigating such data-driven and topology-aware regularizers for optimal power flow and unit commitment problems.
	\bibliographystyle{IEEEtranS}
	\bibliography{power,refs}

\end{document}